# Quantum interference of biphotons at a blazed grating

**Martin Ostermeyer, Dirk Puhlmann, Dietmar Korn**

*University of Potsdam, Institute of Physics and Astronomy, Karl-Liebknecht-Str. 24/25, 14476 Potsdam-Golm, Germany*

[*]*Corresponding author: oster@uni-potsdam.de*

**Abstract:** Correlations between photons are interesting for a number of applications and concepts in metrology in particular for resolution improvements in different methods of quantum imaging. Since Fock-states of N-photons of wavelength $\lambda$ in interference schemes acquire N-times the phase shift of single photons these states can appear as if they had a de Broglie wavelength of $\lambda/N$. A biphoton beam diffracted at a blazed grating shows this reduced de Broglie wavelength. This experiment can be seen as a purification of biphotons of a certain correlation strength on the one hand. On the other hand the evaluation of the one-photon and two-photon rate distributions in the Fraunhofer far field of the grating allows for an analysis of the spatial correlation between the photons. An experimental demonstration of these ideas tested for a biphoton beam generated by parametric down conversion (PDC) is reported. We demonstrate in addition that the existence of higher order spatial modes is important to observe strong spatial correlations and to observe the photonic de Broglie wavelength of $\lambda/2$ for biphotons.

## 1. Introduction

Fock-states of N photons of wavelength $\lambda$ can appear as if they had a de Broglie wavelength [1] of $\lambda_B = \lambda/N$. Examples where light constituent of Fock-states behaves as if it had a wavelength divided by the photon number have been investigated e.g. for the diffraction at ordinary gratings [2, 3, 4], in quantum lithography [5] or related interference schemes [6, 7, 8, 9], and for nonlocal aspects of the de Broglie wavelength of Fock states [10]. Such a reduced wavelength can be interesting for an improved spatial resolution in different concepts of advanced imaging methods in the area of quantum imaging [11, 12]. In addition, nonclassical spatial correlations between the photons can play a key role [13] for these advanced imaging methods. In this paper we report on the characterization of the spatial correlation strength of biphotons from parametric down conversion by Fraunhofer far field distributions behind a blazed grating. Given a certain strength of the spatial photon correlations these distributions show orders of diffraction that relate to a photonic de-Broglie wavelength of half the original wavelength of the photons. We illustrate the connection between the degree of correlation and the spatial multi-mode character of the light.

Our investigations connects to the investigations of Shimizu et. al. [4, 14] where diffraction of biphotons at an ordinary grating was considered. In the course of our characterization within this paper we present the prospects to isolate biphotons of a certain spatial correlation strength so that they can be discriminated from less correlated photon pairs. This separation of biphotons will be enabled by the special properties of a blazed grating in contrast to an ordinary grating. For a certain wavelength called blazed wavelength an ideal blazed grating has only one order of diffraction (see e.g. [15]), the first order.

## 2. Calculating the single and two-photon rate of diffraction behind the blazed grating

Fig. 1 shows the diffraction efficiency in the first order of diffraction (OD) of our blazed grating. It has a grating period of 25 μm and is manufactured for a blaze wavelength of 500 nm.

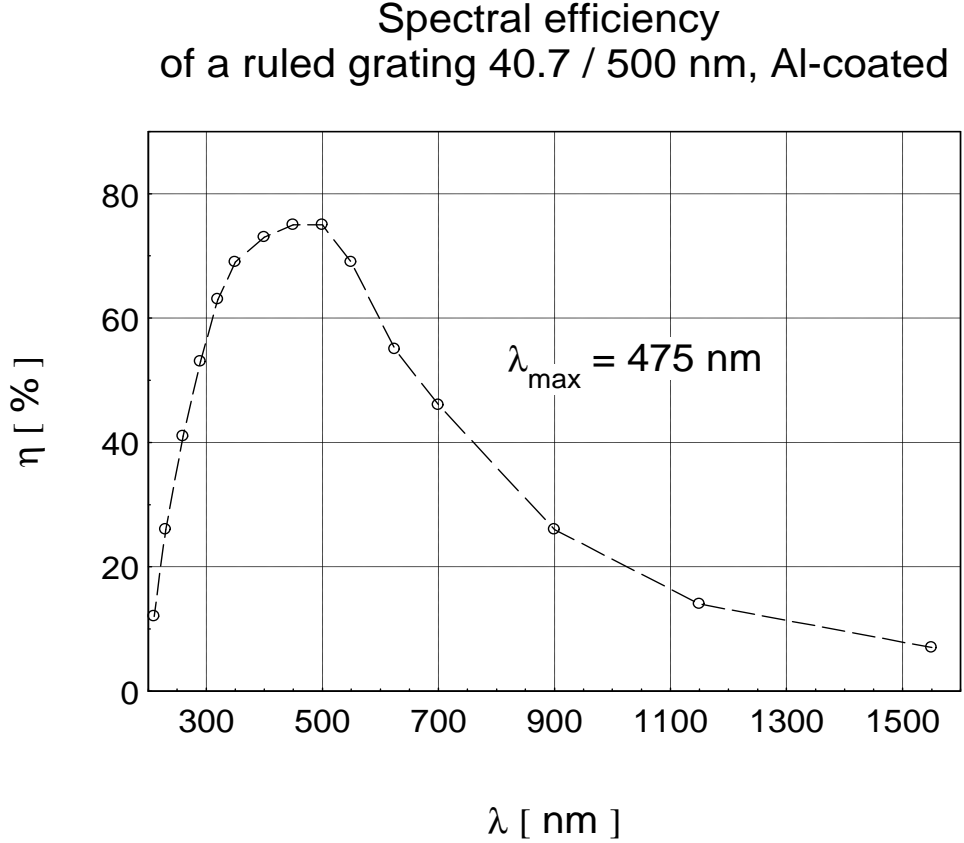

*Fig. 1: Diffraction efficiency of our blazed grating for the first OD for different wavelengths.*

For the calculation of the diffraction at the blazed grating we utilize a semi-classical model as it was applied by Shimizu et. al [14] to the diffraction of biphotons at ordinary gratings before. The two photon state at the near field plane with transverse coordinates $x_1$ and $x_2$ is formulated by:

$$|\psi\rangle = \int dx_1 \int dx_2 F(x_1, x_2) \hat{a}^\dagger(x_1) \hat{a}^\dagger(x_2) |0\rangle$$

$\hat{a}^\dagger(x)$ is the creation operator of a photon at position x. $F(x_1, x_2)$ is the two-photon probability amplitude at the diffracting object with a correlation function $G(x_1-x_2)$ between the positions of the photons:

$$F(x_1, x_2) = \frac{1}{2}[A(x_1)A(x_2) + A(x_1)A(x_2)]G(x_1 - x_2)$$

A(x) is the probability amplitude of single photons at position x which will be used to express the transmission function of the blazed grating. Since both photons pass through the same object which is in our consideration the blazed grating the probability functions for both photons are identical. The entire function F has to be symmetric since it describes the probability amplitude of bosons. The two-photon count rate in the near field of the diffracting object is given by

$$R^{(2)}(x_1, x_2) = \langle\psi|\hat{a}^\dagger(x_2)\hat{a}^\dagger(x_1)\hat{a}(x_1)\hat{a}(x_2)|\psi\rangle = |\langle 0|\hat{a}(x_1)\hat{a}(x_2)|\psi\rangle|^2 = |F(x_1, x_2)|^2$$

The two-photon count rate in the Fraunhofer far field behind the grating is calculated in the same manner:

$$R^{(2)}(k_{\perp 1}, k_{\perp 2}) = |F(k_{\perp 1}, k_{\perp 2})|^2$$

$k_{\perp 1,2}$ are the transverse components of the wave vector of the two photons. The annihilation operators of the far field are calculated from the near field operators by Fourier transformation. The single photon count rate can be calculated by integrating over all undetected photons at $k_{\perp 2}$ [18]:

$$R^{(1)}(k_{\perp 1}) = \int dk_{\perp 2} R^{(2)}(k_{\perp 1}, k_{\perp 2})$$

Photons generated from parametric down conversion are correlated with respect to their near field position which is the position exactly in the nonlinear crystal plane (see e.g. [21]). They are anti-correlated with regard to their transverse momentums which are equivalent to their far field positions. To characterize the strength of the spatial correlation of our photons we use a phenomenological correlation function G where the near field case corresponds to $x_1 - x_2$ and the far field case corresponds to $x_1 + x_2$ with a correlation width $\sigma_{corr}$:

$$G(x_1, x_2) = \exp\left[-\frac{(x_1 \pm x_2)^2}{2\sigma_{corr}^2}\right]$$

We first consider the ramifications in the near field case. Limiting cases of such a correlation between the photons are given by the cases without any correlation $G(x_1, x_2)$ = const and perfect spatial correlation expressed by $G(x_1, x_2) = \delta(x_1-x_2)$. In the case without any correlation the two-photon amplitude becomes separable in functions of $k_{1,\perp}$ and $k_{2,\perp}$. So for equal momenta in the far field $k_{\perp 1} = k_{\perp 2} = k_\perp$ the two-photon count rate and single photon count rate are simply expressed by different powers of the Fourier transform **F**[A] of the single photon amplitude A:

$$R^{(2)}(k_\perp, k_\perp) \propto |\mathbf{F}[A](k_\perp)|^4$$
$$R^{(1)}(k_\perp) = |\mathbf{F}[A](k_\perp)|^2$$

This basically means that single and two-photon count rate look qualitatively the same. They are just different by the square. On the other hand, in case of perfect correlation separability is lost and calculating single and two-photon count rate gives [14]

$$R^{(2)}(k_\perp, k_\perp) \propto \left| F[A](2k_\perp) \right|^4$$
$$R^{(1)}(k_\perp) = const.$$

The Fraunhofer far field distribution of the two photon count rate appears as if it stems from light with half the wavelength of the photons. This means in case of our blazed grating that there is only one strong order of diffraction (OD) but at the position where it was expected for light of half the wavelength. The single photon count rate becomes constant, no interference contrast remains. Apparently the biphotons interfere as if they were one particle that had a de-Broglie wavelength of half the original wavelength. This can be interpreted by the duality between partial coherence and partial entanglement (see e.g. [19]). The biphoton wave amplitude is mathematically analogous to the second order correlation function for the electrical field (which would be the first order correlation for the intensity). Non-separability in the biphoton wave amplitude is associated with entanglement. Non-separability of the second order correlation function on the other hand is connected with loss of coherence and hence loss of interference visibility. Thus, in the limiting case of spatially entangled photons the single photons totally lose their spatial coherence. Thus, interference contrast in the single photon count rate of the Fraunhofer far field behind the grating vanishes.

To illustrate the effect of biphoton diffraction at a blazed grating we calculate three diffraction cases for different spatial correlation strength between the photons of a biphoton, a weakly correlated case with $\sigma_{corr} = 100$ µm, a medium correlated case with $\sigma_{corr} = 9$ µm and a strongly correlated case with $\sigma_{corr} = 0.1$ µm. Whether the correlation appears as relatively strong or relatively weak in our experiments depends on the relation of $\sigma_{corr}$ to the grating period d. If $\sigma_{corr}$ is smaller than the grating period d the correlation would be called relatively strong and both photons can pass in principle through one grating period as biphoton provided the crystal is imaged onto the grating. The calculations are carried out for exactly the grating used in the experiment reported on below with two different spot diameters on the grating. The distributions are shown in dependence of diffraction angle $q = k_\perp/k$ for the direct comparison with our measurements. The first examples shown in Fig. 2 are calculated for an illumination of several grating periods with a spot diameter of 100 µm.

The three diagrams in one row show to the left the two-photon rate as contour plot with the position of two (single photon) detectors at positions $k_{\perp 1}$ and $k_{\perp 2}$. The diagram in the middle shows the evaluation of a cut through the contour plot for $k_{\perp 1} = k_{\perp 2}$ for the two-photon rate. Whereas the diagram to the right shows the single photon rate $R^{(1)}$. The top row of Fig. 2 shows the results in case of almost uncorrelated photons ($\sigma_{corr} = 100$ µm). As expected from the explanations given above for the uncorrelated case two- and single-photon rate just differ by the square. There is one strong $1^{st}$-OD expected for a wavelength of 780 nm (red) and a remaining part because of the limited diffraction efficiency in the $0^{th}$-OD.

For strongly correlated photons (Fig. 2, bottom) one strong OD at the position expected for a $1^{st}$-OD of light of a wavelength of 390 nm (blue) is found whereas the remaining $1^{st}$-OD expected for red light (780 nm) is comparably faint. As expected the interference contrast in the single photon rate is almost zero. The ratio of heights of the red and blue OD changes continuously with the strength of the spatial correlation between the photons of a biphoton. The middle of Fig. 2 shows a calculation for medium correlation strength. The interference contrast of the single photon rate changes in anti-correlation with the relative height of the blue OD. Because the diffraction efficiency of the grating does not reach 100 % we get some photons in the 0$^{th}$-order and the 3$^{rd}$ OD in the blue. The second order of the blue OD falls together with the 1$^{st}$ OD in the red so that they cannot be distinguished.

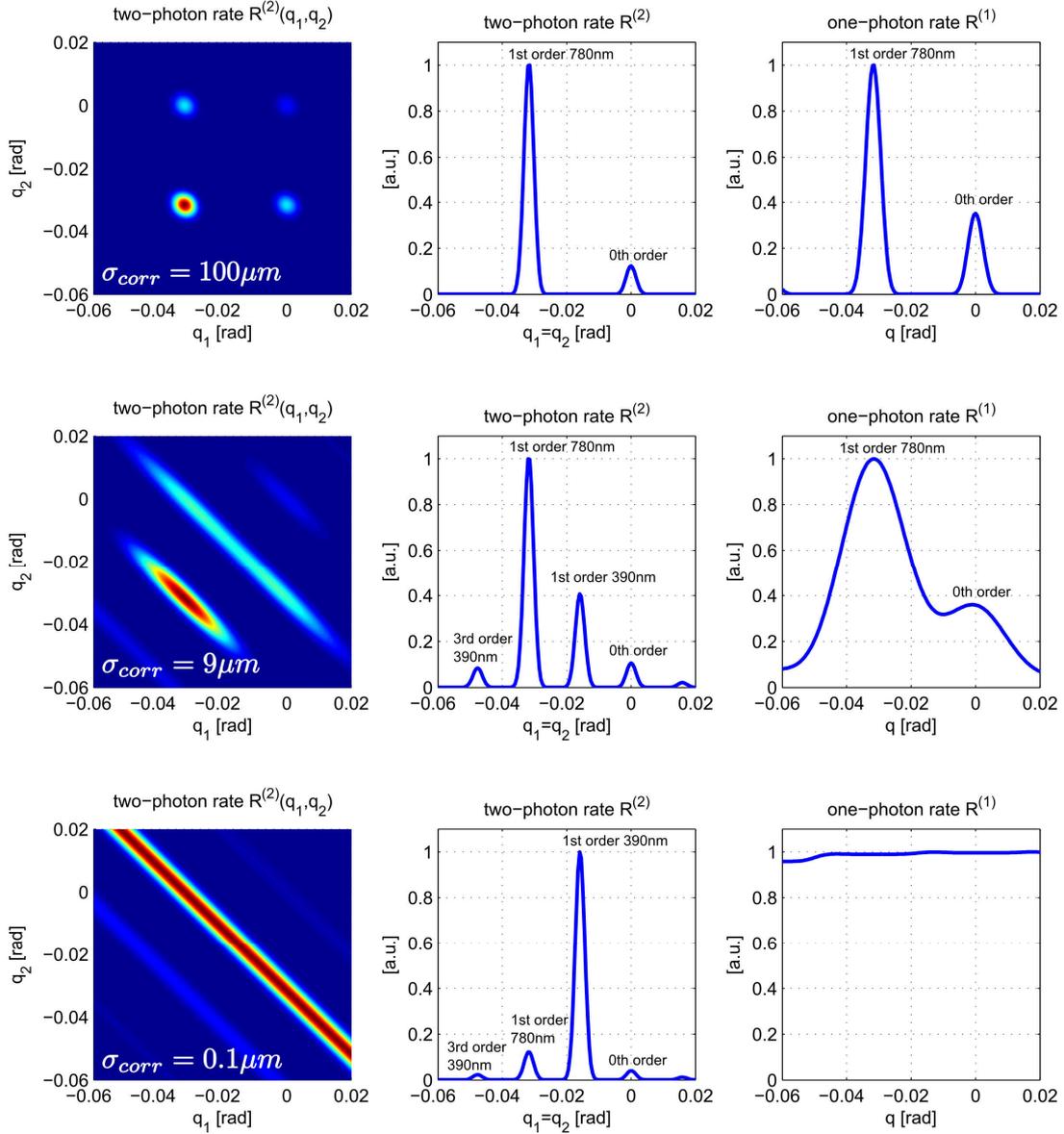

*Fig. 2: Calculated photon rates for Fraunhofer diffraction behind the blazed grating in near field illumination with a spot diameter of 100 µm on the grating for different spatial correlation strength between the photons of a pair. Top: $\sigma_{corr} = 100$ µm, middle: $\sigma_{corr} = 9$ µm, bottom: $\sigma_{corr} = 0.1$ µm*

The absolute correlation width is expected to scale with the diameter of the single photon spatial mode size in the imaged near field plane of the nonlinear optical crystal where the photons are generated (see section 3). In our experiment we choose smaller spot sizes on the grating to obtain smaller absolute correlation width compared to the grating period. On the other hand If fewer grating periods are illuminated the orders of diffraction (OD) become wider (see Fig. 3, top). In further anticipation of our experiment we also calculate the single and two-photon rate for the case of limited angular resolution of the detectors in the Fraunhofer far field of the grating of

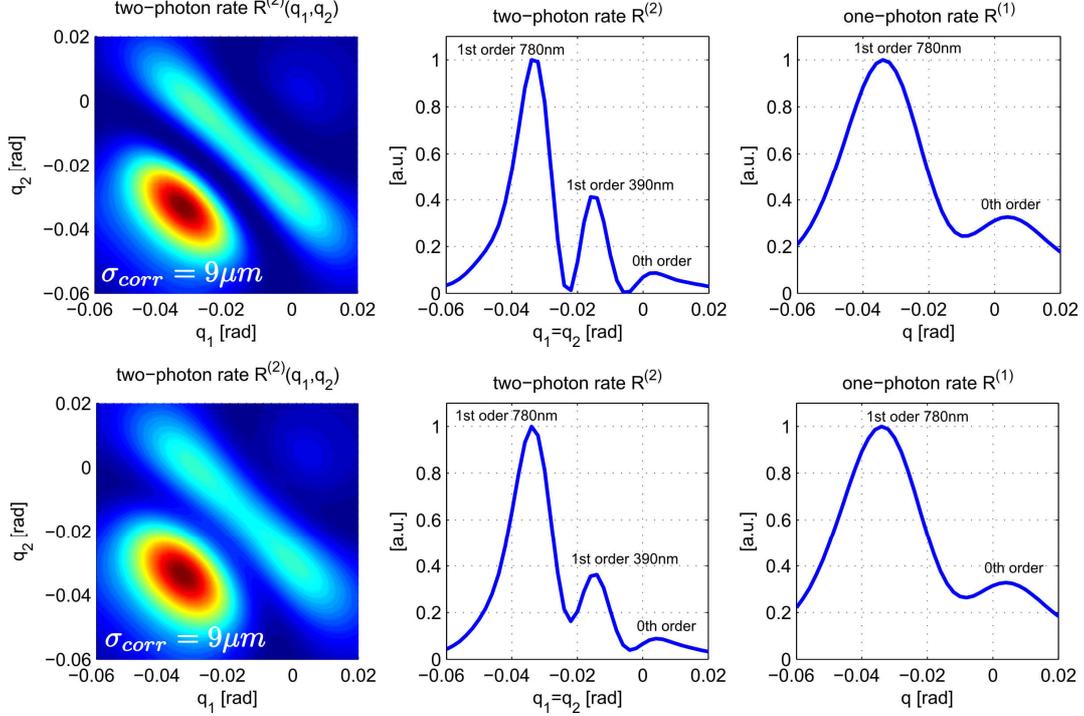

*Fig. 3: Calculated photon rates for Fraunhofer diffraction behind the blazed grating in near field illumination with a spot diameter of 30 µm for a correlation width between the photons of a pair of $\sigma_{corr}$ = 9 µm. Top: angular resolution 0.5 mrad, bottom: angular resolution 10 mrad*

10 mrad. As expected this causes the contrast in the distributions to decrease somewhat (see Fig. 3, bottom).

In far field illumination of the grating the two photons of a strongly correlated biphoton "do not pass through one common slit". Consequently they do not produce any interference contrast in the two photon rate for the profile cut of $q_1 = q_2$ (see Fig. 4 bottom). The smaller the degree of spatial correlation the more the distributions of the count rates of the Fraunhofer diffraction behind the blazed grating in near and far field illumination assimilate each other as expected (see Fig. 4 top).

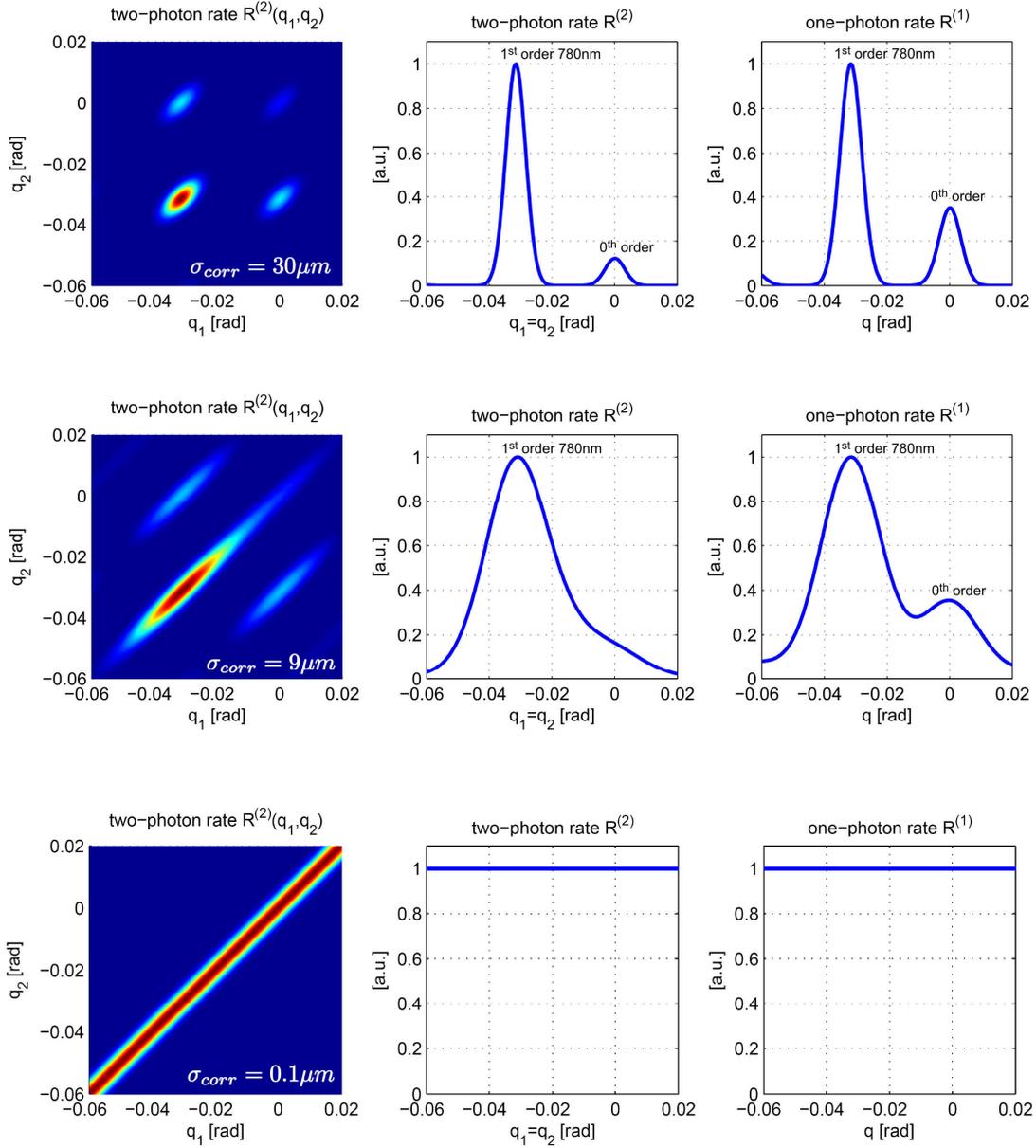

*Fig. 4: Calculated photon rates for Fraunhofer diffraction behind the blazed grating in far field illumination for different spatial correlation strength between the photons of a pair. Top: σ$_{corr}$ = 30 µm, middle: σ$_{corr}$ = 9 µm, bottom: σ$_{corr}$ = 0.1 µm*

## 3. Setup to build a biphoton beam

We generate correlated photons from non-collinear type II parametric down-conversion [16] in a BBO crystal. The schematic of the setup is shown in Fig. 5. The BBO crystal is pumped by a frequency doubled mode locked picoseconds TiSa-laser with a repetition rate of 75 MHz and 20mW average power at a wavelength of 390 nm. The pump spot diameter is 160 µm. The run time differences and walk-off due to the birefringence of the BBO are compensated to first order by a half wave plate and compensation BBO-crystals [16]. By adjusting the compensation crystals we can adjust the two photon Bell-state for $|\psi^+\rangle$ or $|\psi^-\rangle$. For the generation of the biphoton beam the setup is adjusted for the $|\psi^+\rangle$ state. The two photons in $|\psi^+\rangle$-state hit a 50 % beam splitter where by Hong-Ou-Mandel (HOM) interference [17] they reunite to a biphoton beam with probability at the lower and upper port of the beam splitter. We only make use of the biphotons leaving the lower port of the beam splitter. The pairs are detected with a coincidence rate of 10,000 counts/s. We have chosen the non-collinear type II generation process to allow for independent manipulation of the photons before they interfere to produce the biphoton beam. (The full possibility of this independent manipulation is not exploited in the work reported here.) The interference at the beam splitter is adjusted by an optimization of the visibility of the HOM-dip. The photons are detected by single mode fiber-coupled single photon detectors (SPCM-AQRH 15 by PerkinElmer).

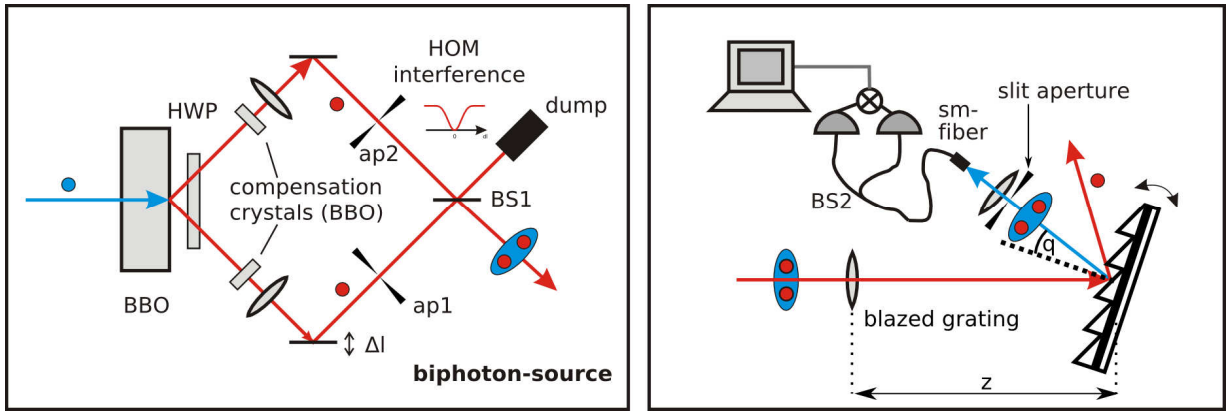

*Fig. 5: Setup to generate a biphoton beam by Hong-Ou-Mandel interference of photons produced from type II non-collinear parametric down conversion (left) and the setup of our grating experiment (right).*

The crystal plane is imaged onto the grating to allow for a near field investigation of the generated photons (compare section 2). This way the strong correlations in the spatial positions of the photons due to its common origin should enable a certain fraction of the photons to pass as a pair through one period of the grating (see section 2). The spot diameter on the grating is 29 µm resulting in a Rayleigh length of around 1 mm. Behind the grating the single photon $R^{(1)}$ and two-photon coincidence rates $R^{(2)}$ are observed in a distance of 10 cm behind the grating. Since this distance is more than one order of magnitude longer compared to the Rayleigh length of the spot this is an observation in the Fraunhofer far field to very good approximation.

The angular resolution of the detection is adjusted by a slit aperture in front of the lens of the fiber coupler. Photons of an angular spread of 10 mrad are coupled into the single mode fiber coupler. At the end of the two ports of the fiber coupler the single and two-photon coincidence count rate is measured by avalanche photo-detectors.

## 4. Experimental observation

The far field distributions in the single and two-photon rate are measured with photon detectors at a fixed position while the grating was turned around an axis perpendicular to the plane of incidence. Since we scanned only a small angular width of around 100 mrad this is equivalent to first order to scanning the detectors in the far field with a fixed grating orientation.

A typical results of such a measurement for the coherence properties of our current biphoton source is shown in Fig. 6. In the measured two-photon count rate two additional pronounced and one weaker OD can be observed. The OD with the highest count rate appears at the angular position of the first OD for the original red wavelength (780 nm) of the photons. In addition there is a distinct OD at the angular position expected for blue light (390 nm). The smaller OD observed to the right is the $0^{th}$-OD. The single photon count rate shows almost no interference contrast since the spatial correlation of the photons is already too strong (see section 0). The measured data is plotted together with the calculated count rates expected for a correlation strength of $\sigma_{corr} = 9$ µm with the angular resolution of our setup of 10 mrad the spot size of 29 µm diameter and 30 mrad full angle. They agree very well.

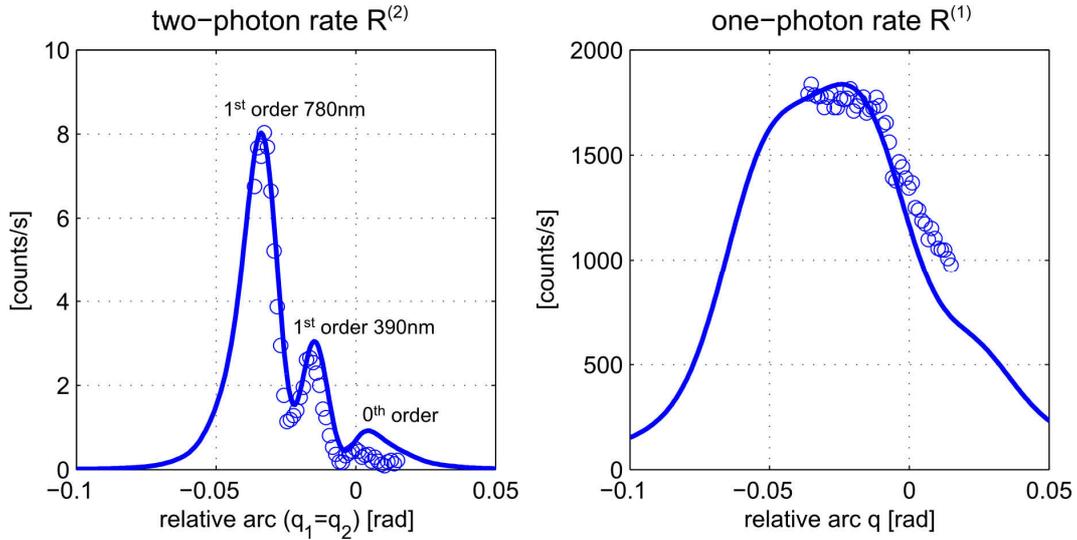

*Fig. 6: Single and two photon rate measured in the Fraunhofer far field behind the blazed grating (circles) and calculated theoretical expectation (straight line) in case of a correlation width of $\sigma_{corr} = 9$µm*

To increase the strength of the spatial correlation the biphoton source and the conditions of propagation in between the source and the grating have to be improved as will be addressed below. The strength of the spatial correlation in the plane of the grating can be decreased by two different methods. First, if the grating is moved away from the plane of the imaged crystal near field the correlations are expected to become weaker. This can be achieved by changing the distance z between the grating and the lens in front of the grating (see Fig. 5). In the experiment the position of the grating has been kept fixed whereas the lens was moved backwards. Second, the correlations of the photons in the near field were weakened by restricting the number of higher order transverse modes which the correlated photons populate. The degree of entanglement or entanglement entropy in such an entangled state of continuous variables is known to grow in a nonlinear fashion with the number of modes that can be populated [20, 22].

The count rates in the Fraunhofer far field measured with the first method are shown in Fig. 7. The correlation strength corresponds to 9 µm, 13 µm and 31 µm respectively as can be taken from the fitted curves. At the lowest level of correlation the blue OD is not visible anymore. This basically means that the photon pairs are not experienced as spatially correlated biphoton by the grating anymore.

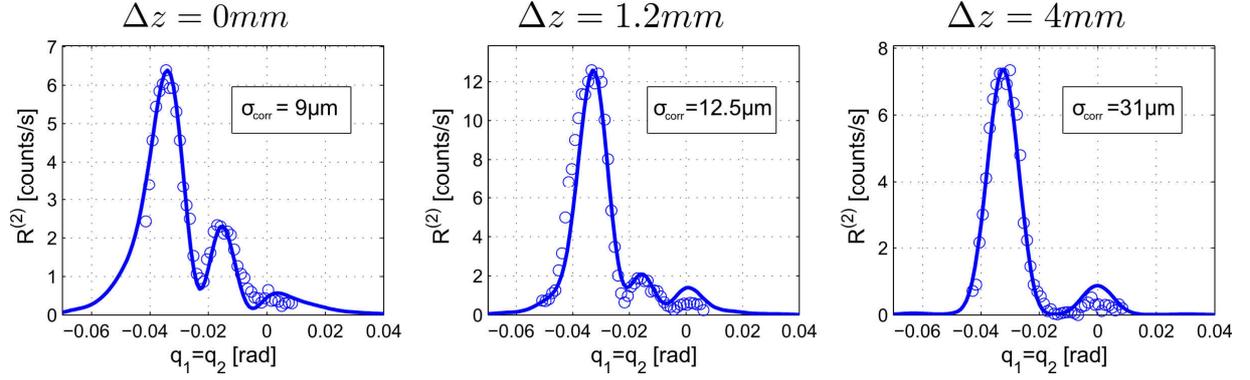

*Fig. 7: Two-photon count rate measured in the Fraunhofer far field behind blazed grating (circles) and calculated theoretical expectation (straight line) for decreasing correlation strength achieved by moving the image plane of crystal near field away from the grating plane.*

The measured count rates in the Fraunhofer far field using the second method are shown in Fig. 8. The restriction of the population of higher order transverse modes for the illumination of the grating was simply achieved by closing the apertures positioned in the far field plane behind the first lens after the BBO crystal (see Fig. 5). Again the 1$^{st}$ OD in the blue disappears if the correlation width of the biphotons becomes too large compared to the grating period. The smallest diameter of the aperture was chosen to be 1.5 mm. The diameter of the transverse fundamental mode that can be expected to be radiated of the pumped area of the BBO-crystal [23] when it was propagated towards the grating was 2.5 mm in the plane of the aperture. The single photon rate gains in contrast along with the vanishing blue order of diffraction and decreasing correlation strength as expected.

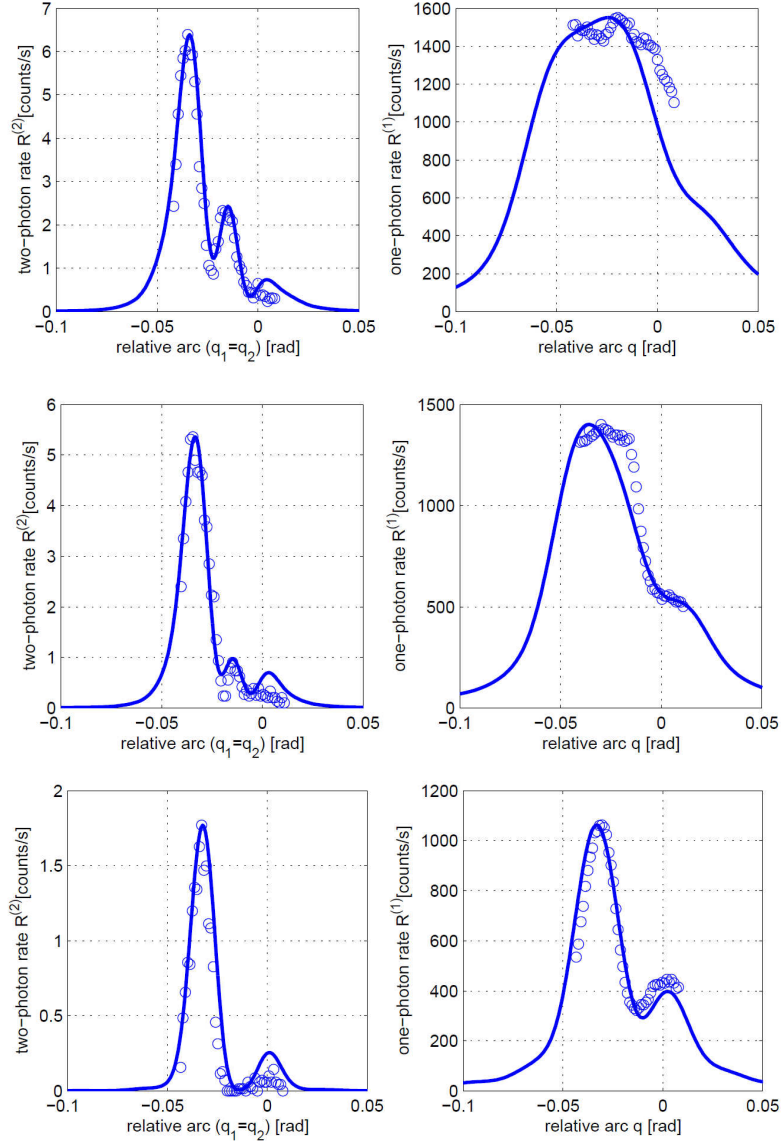

*Fig. 8: Two photon and single photon count rate measured in the Fraunhofer far field behind blazed grating (circles) and calculated theoretical expectation (straight line) for decreasing correlation strength achieved by restriction of higher order transverse modes. aperture diameter top: d=4mm, middle: d=2.5mm, bottom: d=1.5mm*

Stronger diffraction into the blue OD can either be achieved by a wider grating period at equal correlation strength of our source or with increased spatial correlation strength in the grating plane if the grating period stays constant. The correlation strength in our case can be increased mainly by increasing the width of apertures between BBO-crystal and grating plane. An aperture in the crystal plane as might be resulting from a stronger focused pump beam does not influence the correlations in this plane or an imaged near field plane of the crystal. If there are apertures between the crystal plane and the Fraunhofer far field plane the near field correlation strength will be perturbed. This interrelation was used to vary the correlation strength in our experiments (see Fig. 8).

The strength of the observed spatial correlation of the biphotons was mainly limited by the aperture of crystals used for the birefringence compensation that have a diameter of 4 mm 30 cm away from the nonlinear crystal. By using compensation crystals with bigger diameter at a position closer to the generation crystal our correlation could be improved distinctly and hence also the diffraction in the blue could be further increased.

## 5. Conclusion

We realized the diffraction of a biphoton beam at a blazed grating and investigated the consequences of the strength of the spatial correlation between the photons on the Fraunhofer far field distributions of the single and two-photon rates. The blazed grating generates a diffraction pattern that shows only one strong order of diffraction (OD). Diffraction into an OD that is expected for light of half the wavelength of the single photons is observed in the Fraunhofer far field distribution of the two-photon rate if the spatial correlation becomes narrower than the grating period is wide. The ratio of the heights of the orders of diffraction at the wavelength $\lambda$ of the single photon and the biphoton's de Broglie wavelength $\lambda/2$ constitutes a measure of the biphotons spatial correlation strength. Thus, such a diffraction setup can be used to sort or divide biphotons by their correlation strength. The biphotons in OD of $\lambda/2$ could then be used as a purified beam of correlated photons for further applications. A separate spatial correlation characterization of the OD's of $\lambda$ and $\lambda/2$ is envisaged.